# Visualizing Microwave-Driven Dynamics of Antiskyrmions and Surface Skyrmions

Zhuolin Li[1*#], Daisuke Nakamura[1*], Spencer Reisbick[2*], Chuhang Liu[2,3], Myung-Geun Han[2], Kosuke Karube[1], Wataru Koshibae[1], Naoto Nagaosa[1,4], Yasujiro Taguchi[1,5], Yimei Zhu[2,3#], Yoshinori Tokura[1,6,7], Xiuzhen Yu[1,8#]

[1]RIKEN Center for Emergent Matter Science (CEMS), Wako, Japan

[2]Condensed Matter Physics and Materials Science Division, Brookhaven National Laboratory, Upton, NY, USA

[3]Department of Physics and Astronomy, Stony Brook University, Stony Brook, NY, USA

[4]Fundamental Quantum Science Program, TRIP Headquarters, RIKEN, Wako, Japan

[5]Baton Zone Program, TRIP Headquarters, RIKEN, Wako 351-0198, Japan

[6]Department of Applied Physics, The University of Tokyo, Tokyo, Japan

[7]Tokyo College, The University of Tokyo, Tokyo, Japan

[8]Institute of Science Tokyo, Tokyo, Japan

＊contributed equally

# corresponding authors: zhuolin.li@riken.jp, zhu@bnl.gov, yu_x@riken.jp

## Abstract

Microwaves provide coherent access to low-energy excitations and serve as effective probes of high-frequency spin dynamics in quantum and magnetic systems. For topological spin textures, microwave excitation is expected to generate rich collective responses, yet direct real-space observation of ultrafast dynamics remains limited. Here we use time-resolved Lorentz transmission electron microscopy to visualize microwave-driven dynamics in a hybrid antiskyrmion structure composed of a central antiskyrmion and surface skyrmions. We resolve the picosecond evolution of antiskyrmion area and second-harmonic signals, evidencing nonlinear responses of spin textures under microwave excitations. We track the core motions of the antiskyrmion and surface skyrmions, which follow distinct trajectories while sharing the same rotational sense. Micromagnetic simulations reproduce the key observations and associate the dynamic modes with the spatial modulation of the core profile along the thickness. These achievements establish ultrafast electron microscopy as a powerful real-space approach for probing high-frequency microwave-driven dynamics of topological magnetic solitons.

# Introduction

Microwave fields play a critical role in frontier science and technology owing to their ability to provide a high degree of coherent control over diverse low-energy processes[1-3]. This relevance is particularly evident in quantum science, where the characteristic energy scales of leading platforms fall in the gigahertz regime[1,4-6]. Recent advances in quantum transduction[7,8] across disparate excitations and in quantum-enhanced sensing[9-11] have further established microwaves as an essential resource for coherent coupling and hybridization between distinct quantum degrees of freedom. In magnetic systems, microwave fields couple naturally to magnons[12-14], the quasiparticles correlated with spin-wave excitations, thereby providing direct access to collective spin dynamics and enabling strong-coupling quantum hybridization[15-17]. Furthermore, the interaction between complex spin textures and microwave excitation leads to abundant unconventional dynamical responses such as breathing or gyrotropic modes[18-22], opening an additional route for exploring the intrinsic interplay between high-frequency stimuli and topological spin orders.

Magnetic skyrmions, characterized by their noncollinear spin configurations, occupy an important position in microwave studies through their nontrivial topology and characteristic magnon modes[23-25]. The topological charge $Q$ of skyrmions is defined as

$$Q = \frac{1}{4\pi} \iint \mathbf{m}(\mathbf{r}) \cdot \left( \partial_x \mathbf{m}(\mathbf{r}) \times \partial_y \mathbf{m}(\mathbf{r}) \right) d^2\mathbf{r} \tag{1}$$

where $m(r)$ is the unit vector of magnetization. The theoretical and spectroscopic studies have revealed extensive collective modes of skyrmions under microwave excitation[18,19,21,22,26], while direct real-space imaging of these dynamics remains largely unexplored. In contrast, antiskyrmions carry the opposite topological charge $Q$ with reduced rotational symmetry[27-29], indicating dynamical behavior distinct from that of skyrmions. Recent theoretical and experimental studies have demonstrated the unique fourfold azimuthal spin-wave modes in antiskyrmion systems[20,30], making antiskyrmion systems attractive candidates for exploring intricate couplings among magnons, topology and spin-texture symmetry. However, microwave-driven dynamics of topological spin textures, such as antiskyrmions, remain elusive in direct real-space observation.

The atomic spatial resolution and broad experimental versatility of transmission electron microscopy (TEM) make it indispensable in modern physical research[31]. To solve the temporal limitation of conventional TEM, ultrafast TEM (UEM) was developed by integrating pulsed pump–probe schemes into a traditional microscope, extending the temporal resolution into the femtosecond scale[32-34]. This capability has enabled broad applications in condensed-matter and quantum research. Recent developments in radio-frequency (RF) controlled stroboscopic UEM have further enabled direct real-space imaging of spin-wave dynamics[35]. In particular, the laser-free implementation of this technique preserves the intrinsic beam quality of the field-emission electron source, maintaining the performance of the original TEM[35-37], thereby making microwave-driven magnetic phenomena experimentally accessible.

In this work, we employ time-resolved Lorentz TEM (L-TEM) to directly visualize the microwave-driven dynamics of a hybrid antiskyrmion structure with both the central antiskyrmion and the surface skyrmions. We quantify the microwave-induced area change and core oscillation with few-picosecond resolution. The measurements reveal a second-harmonic response, together with distinct core trajectories but consistent rotational sense of the central antiskyrmion and the surface skyrmions. Micromagnetic simulations reproduce the essential experimental features and support a three-dimensional hybrid spin texture in which the core geometry shapes the dynamical entity under microwave excitation.

## Experimental setup for time-resolved measurements

Figure 1 presents the UEM and the real-space magnetic textures investigated in this work. Unlike laser-based UEM, the present measurements were carried out using an RF-controlled stroboscopic UEM scheme, in which the electron beam was temporally gated into pulsed packets by using microwave excitation across an RF strip-line to sweep the electron beam across a chopping aperture at high frequencies (Fig. 1**a**)[35-37]. The RF used to pulse the electrons was then routed through a phase shifter and a frequency doubler to a sample-mounted chip via in-plane waveguides. By tuning the phase shifter, the delay time between the electron pulses and the microwave excitation was controlled, creating stroboscopic snapshots of the skyrmion and antiskyrmion motion at different moments in the excitation cycle (shown in the inset of Fig. 1**a**).

Single-crystalline $(Fe_{0.63}Ni_{0.3}Pd_{0.07})_3P$ (FNPP), which is known to host both skyrmions and antiskyrmions[28,38,39], was used in the current experiment. The lamella specimen was fabricated by a focused ion beam (FIB) system (NB5000, Hitachi) and suspended across the central hole of the chip to maintain electron transparency for TEM imaging. Corresponding scanning electron microscopy (SEM) images and the wiring arrangement of the sample holder are provided in Extended Data Fig. 1. When the microwave signal was introduced into the waveguide, an alternating magnetic field was generated from the waveguide and drove the magnetic textures in the lamella. Owing to the finite tilt (a few degrees) of the mounted lamella relative to the chip plane, the alternating magnetic field at the specimen contained both in-plane and out-of-plane components. In this condition, both breathing and gyrotropic modes are expected to be accessible.

Fig. 1**b** depicts the image of the antiskyrmion textures in the Fresnel mode of L-TEM under a 100-mT field applied normally to the sample plane. Antiskyrmions are embedded in a stripe-domain background and exhibit the characteristic square-like configuration arising from the interplay between anisotropic Dzyaloshinskii–Moriya interaction and dipolar interaction. In addition to the square outline, each antiskyrmion exhibits a core vortex, as shown by the dark circular contrast. The in-plane magnetic induction maps in Fig. 1**c** demonstrate both a square configuration and an inner vortex-like domain.

Previous electron tomography and micromagnetic studies have shown that, near the sample surfaces, the spin configurations of an antiskyrmion can evolve into Neel-type skyrmions due to the demagnetization energy and $S_4$-symmetry breaking[40,41]. The projected magnetic induction from spin-whirl cores near the top and bottom surfaces jointly gives rise to the additional inner vortex-like contrast observed in L-TEM[28]. Figure 1**d** uses the simulated spin texture to demonstrate the *x*-*z* cross-section marked by the grey plane in Fig. 1**c**, illustrating the evolution along the *z* direction from an antiskyrmion in the central region to skyrmions near the surfaces. Figure 1**e** further illustrates the spin configurations at the top surface and in the central region. On this basis, we examine the microwave-driven dynamics of this composite spin texture to provide evidence for possible coupled responses of these two topological spin textures under microwave excitation.

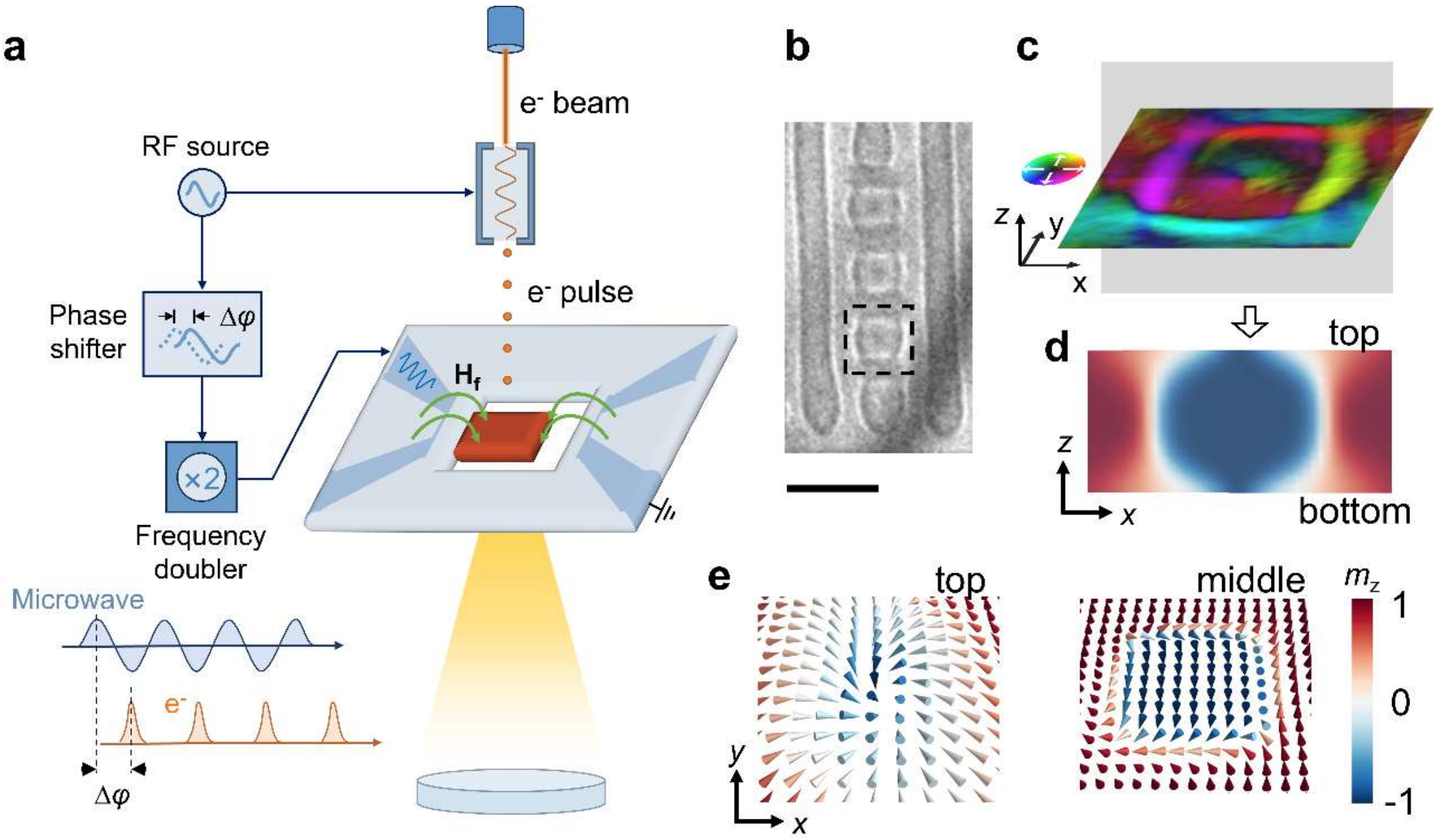

**Fig. 1 | Schematic of the experimental setup and real-space identification of a hybrid antiskyrmion texture. a,** Schematic of the laser-free pulser ultrafast TEM (UEM) system. The electron beam is temporally gated into pulsed electron packets by RF cavities during operation. The RF used to pulse the electron beam is routed through a phase shifter, then frequency doubled before arriving at the chip, enabling phase-locked pump–probe timing control. The green arrow denotes the alternating magnetic field $\mathbf{H}_f$ generated by the in-plane waveguide (dark blue) under microwave excitation. The red lamella suspended across the central hole of the chip depicts the FNPP sample fabricated by focused ion beam (FIB). The inset demonstrates the phase-locked configuration between the pulsed electron packets and microwave. (throughout the manuscript,

time-delay and phase-shift $\Delta\varphi$ are used interchangeably where $\Delta\varphi = 1^\circ \sim 2.7$ ps). **b**, Real-space image of antiskyrmions embedded in a stripe-domain background captured by the Fresnel mode in L-TEM. **c**, In-plane ($x$-$y$ plane) magnetic induction maps of the antiskyrmion labeled by the dashed box in **b**, obtained by transport-of-intensity equation (TIE) analysis[42], with square antiskyrmion frame and inner vortex-like structure. The inset shows the color wheel. **d**, Simulated structure of the hybrid antiskyrmion structure in the $x$-$z$ cross section taken at the position marked by the grey plane in **c**. **e**, The illustrations of the skyrmion at the top surface and antiskyrmion near the central region, respectively. The inset color bar refers to the value of $m_z$. The scale bar is 400 nm.

## Deformation of antiskyrmions under microwave excitation

Microwave excitation at 4 GHz was subsequently introduced to the waveguide and the spectral measurements confirmed that unwanted components of the input signal were suppressed (see Supplementary Note 1). Before performing the time-resolved experiment, we first conducted power-dependent experiments in the Fresnel mode of L-TEM. A measurable response to microwave excitation emerged only above a threshold power of -2.13 dBm and strengthened progressively with increasing microwave power, as shown in Extended Data Fig. 2 and Supplementary Video 1. We therefore fixed the microwave power at 20 dBm to investigate the microwave-excited dynamics of topological spin textures. Using a phase-locking system, the spin texture in the sample was probed by electron pulses at the same repetition frequency of 4 GHz. The antiskyrmion dynamics were recorded over two and a half microwave cycles with a delay time step of 13.9 ps, corresponding to a phase increment $\Delta\varphi = 5^\circ$ (as seen in Supplementary Video 2). Figure 2**a** shows one microwave period with five representative delay times (I–V), and the corresponding UEM snapshots with enhanced contrast. Both the rectangular antiskyrmion and the inner vortex-like contrast remain clearly resolved, highlighting the capability of the instrument to visualize ultrafast spin dynamics in real space.

To quantify the antiskyrmion deformation, we used a custom Python script to track the dark contrast associated with the antiskyrmion domain walls at each delay time, as indicated by the orange outlines in Fig. 2**a**. Details of the data acquisition procedure are provided in Supplementary Note 2. The temporal changes in the averaged width, height and enclosed area of the antiskyrmion were then determined and plotted as black points in Fig. 2**b**-**d**. With the pulsed-electron sampling frequency fixed at 4 GHz, the stroboscopic acquisition selectively captures the system responses synchronized with the sampling frequency and its higher harmonics. (see Extended Data Fig. 3 for details). Nevertheless, noise in the data and uncertainty in outline extraction can introduce additional frequency components into the raw traces. To remove the noise, we therefore applied a numerical bandpass filter retaining only the 4 GHz fundamental and 8 GHz second-harmonic components, shown as solid curves in Fig. 2**b**-**d**. The shaded ribbons represent the one-standard-deviation envelopes of the residuals between the raw data and the filtered curves. The filtered results indicate that the width and height of antiskyrmion domain walls both respond dynamically to microwave excitation but follow distinct temporal traces.

Figure 2**e**-**g** shows the amplitudes and phases of the fundamental and second-harmonic components of the averaged width, height and enclosed area, extracted from the bandpass-filtered traces. The second-harmonic component indicates a nonlinear response to microwave excitation. Extended Data Fig. 4 further compares the correlations between area change and the corresponding width and height modulations using the bandpass-filtered data. The fundamental component of the area change is more strongly associated with the height modulation, whereas the second-harmonic component follows the width modulation more closely. The phase information in Fig. 2**e-g** shows the same frequency-dependent contrast, suggesting that the nonlinear microwave response is closely related to the asymmetric confinement potential imposed by the lateral stripe-domain background.

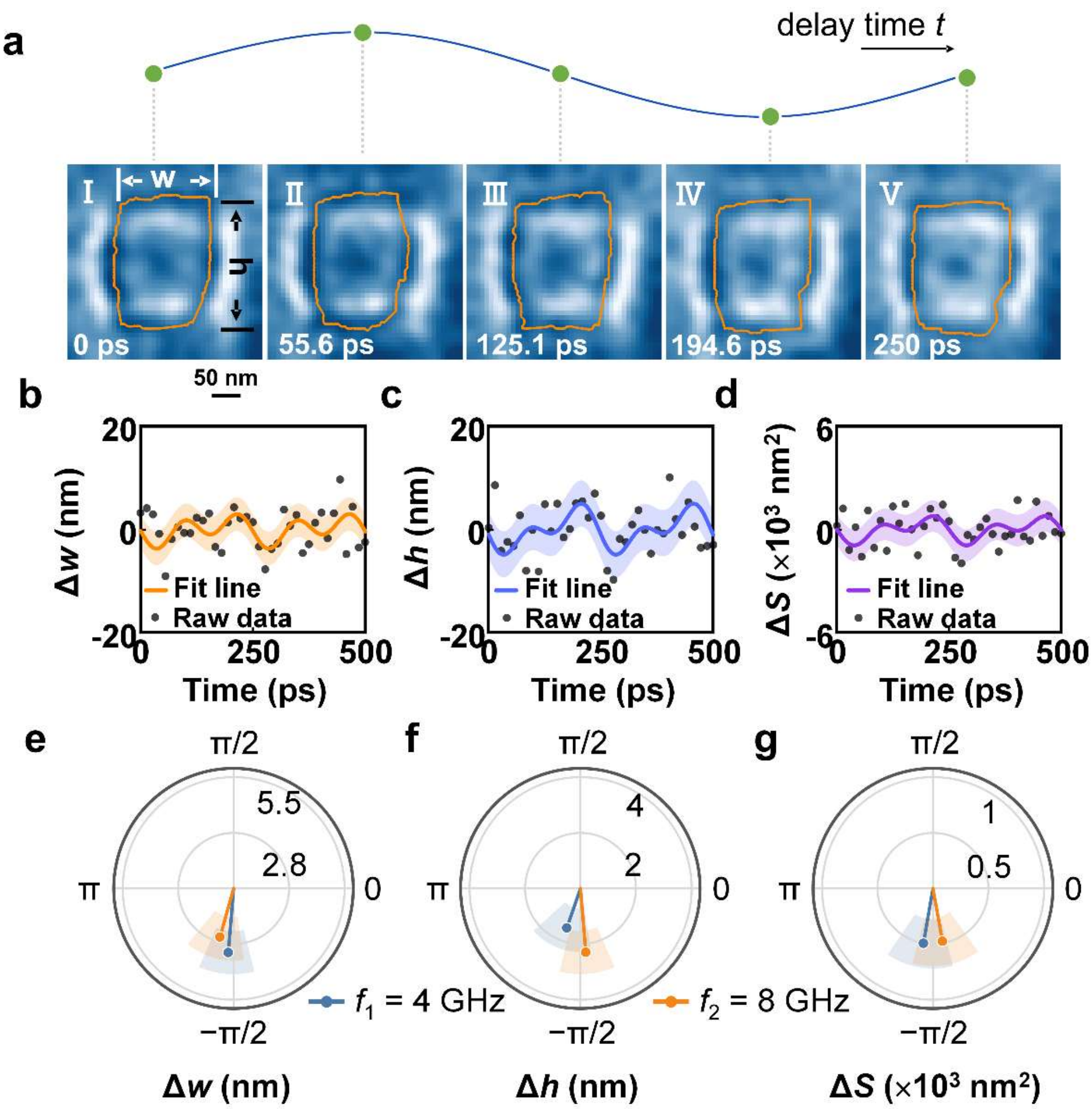

**Fig. 2 | Time-resolved area change of antiskyrmion under 4 GHz microwave excitation. a**, The time-resolved UEM snapshots of an antiskyrmion at five representative delay times (I–V) in

one microwave period. Orange outlines were determined by custom Python script (details in Supplementary Note 2) and indicate the antiskyrmion frame used for quantitative analysis. **b-d**, Temporal evolution of the averaged width change ($\Delta w$, **b**), averaged height change ($\Delta h$, **c**) and the enclosed-area change ($\Delta S$, **d**) defined in **a**. Consecutive frames are acquired with a step of 13.9 ps ($\Delta\varphi = 5°$) set by the pump–probe timing. Black symbols show the experimental data, and solid curves show the corresponding bandpass-filtered traces retaining the 4 GHz and 8 GHz components. The ribbon represents the residual standard-deviation envelope of the raw data around the fitted curves. **e-g**, Amplitudes and phases of the fundamental and second-harmonic components of $\Delta w$ (**e**), $\Delta h$ (**f**) and $\Delta S$ (**g**), extracted from the bandpass-filtered traces. The ribbon represents the residual standard-deviation envelope of the amplitudes and phases. The scale bar in **a** is 50 nm.

## The core dynamics of the hybrid antiskyrmion structures under microwave excitation

Microwave-driven (anti)skyrmion core dynamics are of particular interest, as they are closely linked to the intrinsic topology of the textures. We extracted both the antiskyrmion core and the surface-skyrmion core from the UEM images, with two representative snapshots shown in Figs. 3**a** and 3**b**. In an antiskyrmion without surface skyrmions, the L-TEM contrast appears only at the domain walls and shows no central contrast[40]. We therefore defined the antiskyrmion core as the centroid of the extracted antiskyrmion boundary, marked by the yellow cross. The inner vortex-like contrast originates from the projected superposition of surface skyrmions near the upper and lower surfaces[28,41]; its centroid, marked by a white dot, was therefore used to track the coherent motion of the surface skyrmions. Details of the extraction procedure are provided in the Supplementary Note 2.

The temporal evolution of the core position of the antiskyrmion and surface skyrmion, together with the corresponding bandpass-filtered trajectories, is shown in Fig. 3**c** and **d**. The two types of cores exhibit trajectories with different shapes and amplitudes, indicating that the core dynamics are sensitive to the underlying spin configurations. To quantify their rotational senses, we introduced the signed area, $\chi$, enclosed by each core trajectory,

$$\begin{aligned}\chi &= \frac{1}{2}\oint (xdy - ydx) \\ &= \sum_n \frac{f_n T}{2}\left(\left|C_{\mathrm{CCW},n}\right|^2 - \left|C_{\mathrm{CW},n}\right|^2\right) = \sum_n N_n \chi_n\,,\end{aligned} \tag{2}$$

where $x$ and $y$ denote the coordinates of the core position, $C_{\mathrm{CCW},\,n}$ and $C_{\mathrm{CW},\,n}$ denote the counterclockwise (CCW) and clockwise (CW) components of the $n$th frequency component, respectively. Here, $f_n$ is the frequency of this component, $T$ is the integration time window, $N_n = f_n T/2\pi$ is the corresponding number of cycles, and $\chi_n = \pi\left(\left|C_{\mathrm{CCW,n}}\right|^2 - \left|C_{\mathrm{CW,n}}\right|^2\right)$ represents the signed area per cycle (see details in Methods). Thus, the sign of $\chi$ determines the net rotational sense, whereas its magnitude reflects the effective orbital amplitude, as illustrated in Fig. 3**d**. Here, $\chi > 0$ corresponds to counterclockwise rotation, while $\chi < 0$ corresponds to clockwise rotation.

Figure 3**f** summarizes the signed areas of the antiskyrmion-core and surface-skyrmion-core trajectories in Fig. 3**c** and **d**, calculated using Eq. (2), together with the values for the fundamental and second-harmonic components. The results show that both the antiskyrmion core and the surface-skyrmion core rotate CCW under 4 GHz microwave excitation. This common rotational sense suggests that hybridization between the surface skyrmions and the central antiskyrmion shapes their microwave-driven

response. Both cores also show pronounced second-harmonic components, likely originating from spatial asymmetry and nonsymmetric surrounding confinement.

The microwave frequency was then increased to 5.25 GHz and the corresponding microwave-driven dynamics are provided in Supplementary Video 3. Using the same analysis procedure as that used for the 4 GHz experiment, the corresponding results are shown in Extended Data Fig. 5. Both cores retain CCW rotational motion but exhibit larger orbital amplitudes than those observed in the 4 GHz experiment. This enhanced response suggests that a higher-frequency resonant response may occur near 5.25 GHz.

Taken together, these results reveal rich microwave-driven dynamics in the hybrid antiskyrmion texture. Structural inequivalence gives rise to distinct yet associated responses of the surface skyrmion and the central antiskyrmion. This behaviour identifies microwave excitation as a sensitive probe of internal symmetry and intrinsic coupling in complex spin textures, and establishes UEM as a powerful approach for visualizing time-resolved magnetic dynamics in real space.

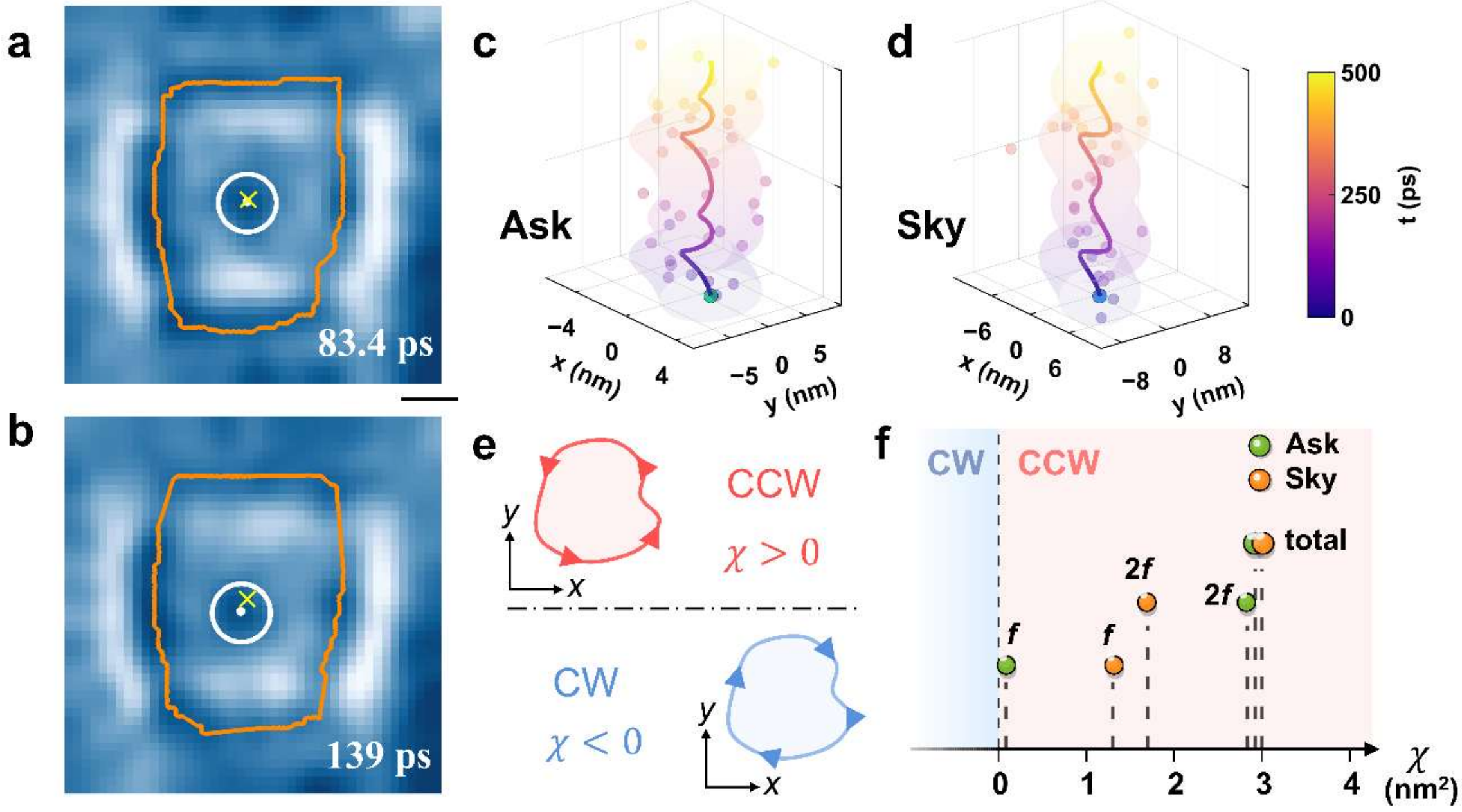

**Fig. 3 | Microwave-driven core dynamics of the central antiskyrmions and the surface skyrmions**. **a**-**b**, Identification of the antiskyrmion core and the surface-skyrmion core at two representative delay times, 83.4 ps ($\Delta\varphi$ = 30°) and 139 ps ($\Delta\varphi$ = 50°). The yellow cross marks the centroid of the extracted antiskyrmion contour, and the white dot marks the centroid of the inner vortex-like contrast, corresponding to the surface skyrmion core. **c**-**d**, The temporal evolution of the extracted core position (color spots) of the antiskyrmion (**c**) and surface skyrmions (**d**) respectively, together with the corresponding bandpass-filtered trajectories. The shaded ribbon denotes the one-standard-deviation envelope derived from the residual covariance of the raw core positions around the fitted trajectory. The color bar encodes the time. **e,** Schematic illustration of the trajectory signed area $\chi$ and the associated sign-dependent rotational sense. **f**, Evaluated $\chi$ for the bandpass-filtered core trajectories of the antiskyrmion and surface skyrmion. The complete trajectories as well as the fundamental and second-harmonic components are evaluated separately. The scale bar in **a** is 50 nm.

## Micromagnetic simulation of microwave-driven hybrid antiskyrmion dynamics

To clarify the origin of the experimentally observed dynamics, we performed micromagnetic simulations using MuMax3[43,44]. A three-dimensional hybrid texture composed of surface skyrmions and central antiskyrmion was constructed (Fig. 4**a**), which has been experimentally confirmed by electron-holographic tomography[41]. Representative spin configurations at selected layers and the simulated L-TEM contrast are shown in Extended Data Fig. 6. The agreement between the simulated and experimental contrast supports the validity of this hybrid structure, providing a reliable basis for interpreting the experimental observations and guiding the subsequent simulations.

A sinc-shaped magnetic-field pulse with broad spectral content was first applied to identify the characteristic resonance modes (see Methods for details). The Fourier spectra of the $m_x$ and $m_y$ components were then calculated, as shown in Extended Data Fig. 7. Multiple response peaks are observed, including peaks at 4.7 and 5.13 GHz (details described in Supplementary Note 3), which are close to the experimentally applied frequencies of 4 and 5.25 GHz.

For direct comparison with the experimental measurements, sinusoidal ac magnetic fields at 4 GHz and 5.25 GHz were applied to the system with both in-plane and out-of-plane components (see Methods). The time evolution of the simulated L-TEM contrast was then recorded, with representative images shown in Fig. 4**b** and Extended Data Fig. 8**a**, respectively. The corresponding movies are provided as Supplementary Videos 4 and 5. The antiskyrmion contour and the vortex-like contrast were tracked using the same procedure as that applied to the experimental data.

Figures 4**c**-**e** show the temporal evolution of the antiskyrmion area and the extracted core positions under 4 GHz excitation, together with bandpass-filtered trajectories. The simulations reproduce the essential experimental features, including the microwave-induced area change and the CCW rotation of the antiskyrmion and surface-skyrmion cores. The insets compare the amplitudes of the fundamental and second-harmonic components extracted from the bandpass-filtered results. The emergence of second-harmonic signals further supports the nonlinear microwave response in the symmetry-broken hybrid texture. The smaller second-harmonic weight in the simulation likely arises from the simplified isolated-texture geometry, whereas the experimental

stripe-domain background provides additional asymmetric confinement that may enhance the nonlinear response. The 5.25 GHz simulations in Extended Data Fig. 8 also show CCW core rotation with a larger amplitude, consistent with the experimental observations. This enhanced response is likely related to the proximity of 5.25 GHz to a resonance peak of the hybrid texture, as indicated by the spectra in Extended Data Fig. 7.

To obtain a more complete picture of the microwave-driven response, we therefore analyzed the simulated core motion in individual $x$-$y$ slices along the thickness. The strongly $z$-dependent signed area of the core trajectory (Fig. 4**f**) indicates distinct rotational responses within the hybrid texture, despite the coherent CCW projected motion observed in L-TEM (Fig. 4**b**). Given that the dynamics of topological spin textures can be influenced by both topology and geometry[20,22,45,46], we then compared the $z$-dependent signed area with the topological charge $Q$ and the normalized second derivative $d^2S_{\mathrm{core}}/dz^2$, where $S_{\mathrm{core}}$ is defined as the area of the core region in each $x$-$y$ slice (Fig. 4**f**). The signed-area profile shows only weak correspondence with $Q$, indicating that the rotational response is not simply determined by the topological charge. Instead, the enhanced signed area correlates more closely with $d^2S_{\mathrm{core}}/dz^2$, suggesting that CCW rotation is supported near regions where the three-dimensional core iso-surface bends strongly along the thickness direction, as illustrated in Fig. 4**g**. This correlation highlights the role of core geometry in shaping the microwave-driven dynamics of the hybrid antiskyrmion texture. Although CW components appear near the surface-skyrmion regions, their phases vary strongly along z and are partly cancelled in projection, whereas CCW components add more coherently (Extended Data Fig. 9). Consequently, the vortex-like contrast observed by L-TEM exhibits a CCW-dominant rotational mode.

The other spectral peaks in Extended Data Fig. 7 were also analyzed and the summarized results are shown in Supplementary Note 3. The resonance frequencies observed here are lower than those reported for microwave-excited antiskyrmion crystals[30]. This difference likely reflects the transition from collective crystal modes to the response of the individual hybrid texture, which is expected to shift the spectrum toward lower frequencies[22,30].

Overall, the micromagnetic simulations reproduce the key experimental observations and resolve the frequency-dependent response of the hybrid texture. These results reveal a complex spectrum of microwave-driven modes in the coupled antiskyrmion–skyrmion structure and establish a close correlation between the resonance behavior and the three-dimensional core geometry of the hybrid texture.

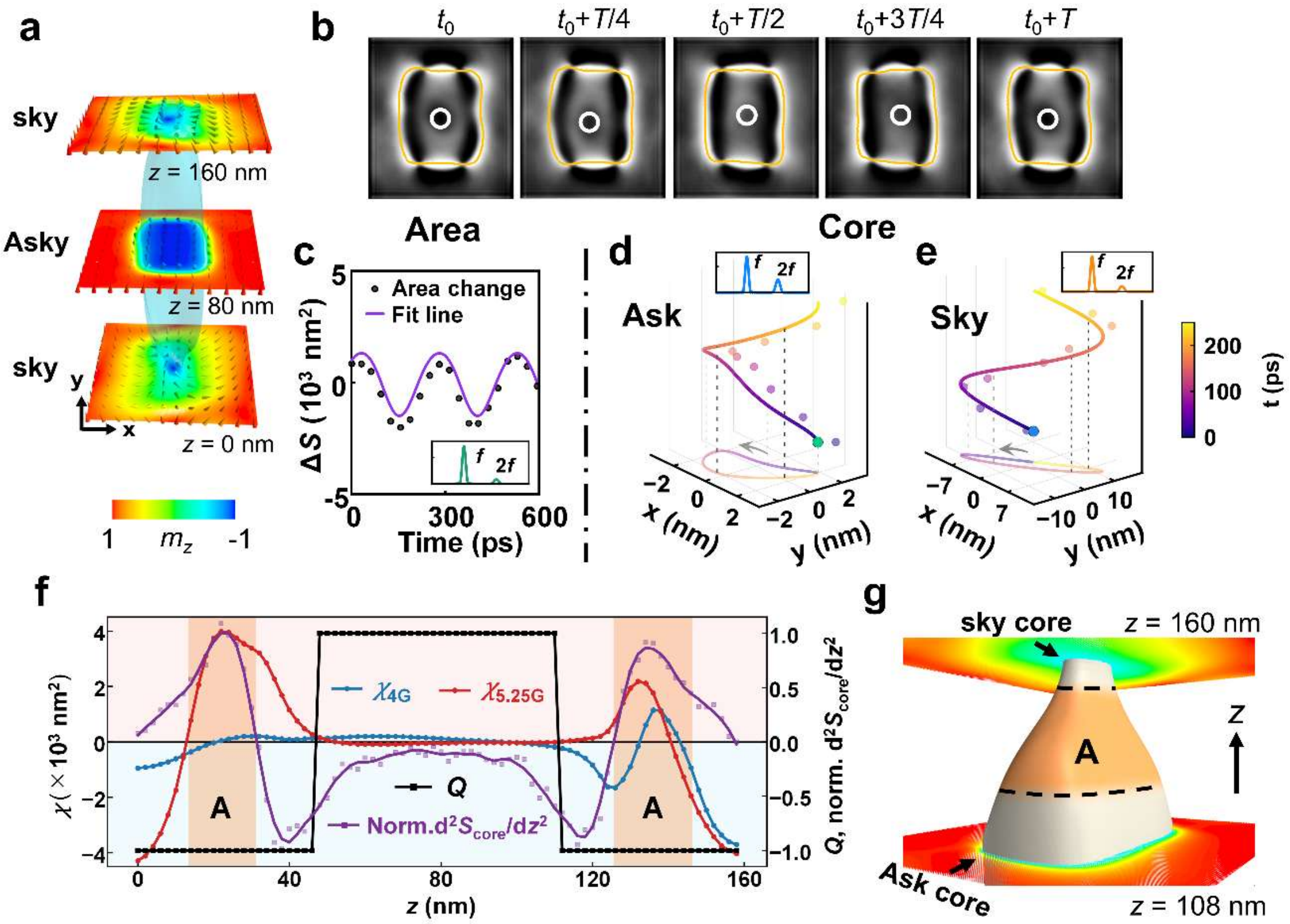

**Fig. 4 | Micromagnetic simulation of microwave-driven dynamics in a hybrid texture. a,** Micromagnetic model of the three-dimensional hybrid texture composed of surface skyrmions

and central antiskyrmions. Representative spin configurations at the top, middle and bottom layers are shown. The semi-transparent iso-surface highlights the connected core region along the thickness direction, defined by $m_z < -0.8$. The color bar represents the out-of-plane magnetization component $m_z$ (red: +1, blue: −1). **b**, Simulated L-TEM contrast images over one period under a 4 GHz sinusoidal magnetic field with both in-plane and out-of-plane components. Yellow outlines mark the extracted antiskyrmion contours, and the white dashed circles indicate the vortex-like contrast associated with the surface-skyrmion core. **c-e,** Extracted temporal evolution of area and core positions from the simulated L-TEM image series. The symbols denote the simulated data, and solid curves denote the bandpass-filtered trajectories containing the fundamental and second-harmonic components. Insets demonstrate relative contributions of the fundamental and second-harmonic components extracted from the bandpass-filtered results. The counterclockwise (CCW) core rotation of the antiskyrmions and surface skyrmions in **d** and **e**, respectively. The color bar encodes the time. **f,** $z$-dependent signed area $\chi$ of the simulated core trajectory, together with the topological charge $Q$ and the second derivative of the core area, $d^2S_{\mathrm{core}}/dz^2$. The enhanced CCW response relates to the regions with large axial curvature, marked by the orange shading defined from the full width at half maximum of the $d^2S_{\mathrm{core}}/dz^2$ peaks. **g,** Schematic illustration of the orange-shaded regions in **f** mapped onto the upper half ($z > 40$ nm) of the three-dimensional core iso-surface.

# Conclusions

In conclusion, we have visualized the microwave-driven dynamics of antiskyrmions in real space using UEM. The temporal imaging reveals microwave-induced change of the antiskyrmion area together with distinct core dynamics involving both the central antiskyrmion and the embedded surface-skyrmion component. The area change exhibits directional anisotropy and the second-harmonic contribution, indicating a nonlinear response associated with the reduced rotational symmetry of the antiskyrmion and the anisotropic confinement imposed by the surrounding stripe-domain background. At the same time, both the antiskyrmion core and the surface-skyrmion core follow different trajectories while sharing a common counterclockwise rotational sense, highlighting the critical roles of the local spin configuration and the coupling between coexisting topological spin textures. Micromagnetic simulations reproduce the coupled dynamical behavior of the hybrid antiskyrmion structure under microwave excitation, revealing a close correlation between the core geometry and the excitation modes. More broadly, this work demonstrates the capability of UEM for direct real-space imaging of microwave-driven spin dynamics in nontrivial topological textures and opens promising avenues for both methodological advances and physical insight into high-frequency spin dynamics.

# Methods

## Sample preparation

The single-crystalline bulk sample of $(Fe_{0.63}Ni_{0.3}Pd_{0.07})_3P$ (FNPP) was synthesized using the self-flux method, employing iron (Fe), nickel (Ni), palladium (Pd), and red phosphorus (P) sealed in an evacuated quartz tube. In-plane Au waveguides were fabricated on Si-based chips by photolithography followed by metal evaporation. For the in-situ L-TEM observations, sample plates were fabricated from FNPP single crystals utilizing a FIB technique. The (001) plane was selected for the sample, with the edges aligned along the <110> direction to align with the spiral wave vector q, optimizing the observation of skyrmion dynamics. The dimensions of the rectangular sample were approximately 12 μm in length, 15 μm in width, and 160 nm in thickness. The fabricated lamella was then transferred in the FIB system to the central opening of the microwave chip and attached to one side of the hole for subsequent UEM measurements.

**The time-resolved imaging in microwave-driven UEM**

Time-resolved measurements were carried out in a customized JEOL JEM-2100F transmission electron microscope operated at 200 kV and equipped with a field-emission gun. A radio-frequency electron pulser was mounted beneath the gun chamber to chop the continuous beam into a pulsed probe while maintaining the high coherence of the original source. In the pulsing stage, a 2 GHz (2.625 GHz) transverse RF field was applied to the first travelling-wave strip line (K1), which swept the electron beam across a 25 μm chopping aperture and generated discrete electron packets. Owing to the bidirectional sweeping process, the effective repetition rate of the pulses reaching the sample was 4 GHz (5.25 GHz). A second strip line (K2) driven at the same frequency, but with a 180° phase shift was used to compensate the angular and momentum dispersion introduced during beam chopping, thereby restoring the trajectory of the pulsed beam and preserving the imaging performance of the microscope.

To realize synchronized pump–probe operation, the RF signal downstream of K1 was directed through isolators to suppress reflections caused by impedance mismatch in the later stages of the circuit. The signal timing and synchronization were controlled with a phase shifter, frequency doubler, and bandpass filter to produce a well-defined excitation waveform whose phase relative to the electron pulses could be tuned precisely. A fraction of the RF output was fed back to a control system to maintain phase locking throughout the measurement. In this way, the relative delay between excitation and probe was controlled through the phase shifter, enabling stroboscopic observation of RF-driven dynamics with few-picosecond timing precision.

Microwave excitation was delivered to the specimen through an RF-compatible TEM holder with a transmission efficiency above 80% in the gigahertz range. The input RF power to the holder was 20 dBm. Time-resolved Lorentz TEM images were recorded with a defocus of 7 mm and an acquisition time of 120 s per frame corresponding to $0.6 \times 10^{12}$ snapshots averaged for each frame. Temporal sampling was achieved by incrementing a delay step of 13.9 ps ($\Delta\varphi = 5°$).

A direct electron detector (Dectris Quadro) was used for all time-resolved TEM measurements to significantly improve the signal-to-noise ratio relative to conventional CCD-based acquisition. Custom control software was also developed to coordinate microscope operation and detector readout, allowing flexible adjustment of acquisition settings such as exposure conditions, thresholding, and image-processing parameters.

## UEM Data Processing

The experimental and simulated UEM image sequences were analyzed using a custom Python workflow that converts stacks of L-TEM frames into contrast-enhanced processing images, structural contours, frame-by-frame displacement tables and frequency-filtered dynamic components. The in-plane core displacement with a complex coordinate was then defined to determine the rotational sense of each core trajectory,

$$z(t) = x(t) + iy(t) = \sum_{n} \left( C_{\mathrm{CCW},n} e^{i f_n t} + C_{\mathrm{CW},n} e^{-i f_n t} \right), \quad (3)$$

where $C_{\mathrm{CCW},n}$ and $C_{\mathrm{CW},n}$ denote the counterclockwise and clockwise circular components of the $n$th frequency component, respectively. Details of the implementation are provided in Supplementary Note 2. To facilitate automated detection, the processing images were rendered with enhanced contrast, as exemplified by the sharpened image in Supplementary Fig. 2**d**. For clearer visualization in Fig. 2**a** and Fig. 3**a**-**b**, the displayed images were rendered with moderately reduced contrast and an alternative colormap, with the antiskyrmion boundary and inner vortex-like contour extracted by the workflow overlaid on the images.

For the dataset acquired under $f$ = 5.25 GHz microwave excitation, the image sequence demonstrates a downward shift of antiskyrmions by approximately 69 nm at a delay time of 152.9 ps ($\Delta\varphi$ = 55°). This displacement is likely associated with a microwave-induced perturbation, which drives the antiskyrmion away from its initial state into another metastable position. In the subsequent analysis, this translational displacement was subtracted to enable a clearer comparison of the core trajectories. The correction procedure is described in Supplementary Note 2.

## Micromagnetic simulation

Micromagnetic simulations were performed using the GPU-based package MuMax3[43,44,47] to model the microwave-driven dynamics of the hybrid antiskyrmion texture. The magnetization dynamics were obtained by numerically solving the Landau–Lifshitz–Gilbert equation including adiabatic and non-adiabatic spin-transfer torque terms,

$$\frac{d\mathbf{M}}{dt} = -\gamma_0 \mathbf{M} \times \mathbf{H}_{\text{eff}} + \frac{\alpha}{M_s}\left(\mathbf{M} \times \frac{d\mathbf{M}}{dt}\right) \quad (4)$$

Here, $\mathbf{M}$ is the magnetization vector, $M_s$ is the saturation magnetization, $t$ is time, $\gamma_0$ is the absolute gyromagnetic ratio. $\alpha$ is the damping parameter.

In the FNPP, Ni, Pd, and P do not contribute appreciably to the magnetic moment at room temperature[28,48]. Instead, the magnetism is dominated by the Fe atoms. The simulations were therefore performed within a single-sublattice micromagnetic framework. The effective field was calculated from the functional derivative of the total magnetic energy density,

$$H_{\text{eff}} = -\mu_0^{-1} \partial\varepsilon / \partial\mathbf{M}, \quad (5)$$

with

$$\varepsilon = A(\nabla\mathbf{m})^2 + D\left[\mathbf{m}\cdot(\hat{\mathbf{x}}\times\partial_x\mathbf{m}) - \boldsymbol{m}\cdot\left(\hat{\mathbf{y}}\times\partial_y\mathbf{m}\right)\right]$$
$$- K_\mathrm{u}(\mathbf{m}\cdot\hat{\mathbf{z}})^2 - \mu_0 M_\mathrm{s}\mathbf{m}\cdot\mathbf{H} - \frac{M_\mathrm{s}}{2}\mathbf{m}\cdot\mathbf{B}_\mathrm{d} \quad (6)$$

where $A$ is the exchange stiffness, $D$ is the anisotropic Dzyaloshinskii-Moriya interaction constant, $K_\mathrm{u}$ is the uniaxial anisotropy constant, $\mathbf{H}$ is the applied magnetic field, $\mathbf{B}_\mathrm{d}$ is the demagnetizing field, and $\mathbf{m} = \mathbf{M}/M_\mathrm{s}$ is the normalized magnetization. Thermal fluctuations were not included, and all calculations were performed at $T$ = 0 K, as thermally activated motion was negligible under the present experimental conditions.

The material parameters were chosen on the basis of previous experimental and theoretical studies[28,39,49]. We used $A$ = 8.1 pJ/m, $K_\mathrm{u}$ = 31 kJ/m$^3$, $M_\mathrm{s}$ = 417 kA/m, $D$ = 0.2 mJ/m$^2$, $\alpha$ = 0.03. In the simulation, $x$ and $y$ were aligned with the crystallographic $[110]$ and $[\bar{1}10]$ directions of the experimental specimen, respectively, to maintain consistency with the measurement geometry. The computational region was discretized into 180 × 200 × 80 cuboids with a mesh size of 2 nm, using periodic boundary conditions.

The external magnetic field used in the simulations was defined as,

$$\mathbf{B}_\mathrm{ext}(t) = B(t)\hat{\mathbf{x}} + [B_\mathrm{s} + B(t)]\hat{\mathbf{z}}, \quad (7)$$

where $B_\mathrm{s}$ is a static perpendicular field of 150 mT applied along the $+z$ direction to stabilize the texture. For pulsed-field spectral analysis, $B(t)$ was set to the sinc-shaped pulse,

$$B(t) = B_\mathrm{sinc}(t) = B_\mathrm{amp}\frac{sin[\pi f_0(t-t_0)]}{\pi f_0(t-t_0)}, \quad (8)$$

where $B_\mathrm{amp}$ = 7 mT and $f_0$ = 50 GHz. This sinc excitation provides broad spectral content from 0 to 25 GHz and thus enables simultaneous excitation of multiple dynamical modes.

For ac-field simulations, $B(t)$ was instead defined as

$$B(t) = B_0\sin(2\pi f t) \quad (9)$$

The amplitude $B_0$ was set to 7 mT, and the frequency $f$ was set to 4 GHz or 5.25 GHz to mimic the microwave excitation used in the experiment. No artificial damping boundary or absorbing potential was introduced.

## Data availability

The main data that support the findings of this study are available within the article and supplementary videos.

## Code availability

The micromagnetic simulator MUMAX$^3$ used in this work is publicly accessible at http://mumax.github.io/index.html. The custom Python scripts used for image processing are available at https://github.com/Leeovo-zl/UEM-workflow.

## References

1 Zuo, X. *et al.* Cavity magnomechanics: From classical to quantum. *New J. Phys.* **26**, 031201 (2024).

2 Yao, J. Microwave photonics. *J. Lightwave Technol.* **27**, 314–335 (2009).

3 Blais, A., Grimsmo, A. L., Girvin, S. M. & Wallraff, A. Circuit quantum electrodynamics. *Rev. Mod. Phys.* **93**, 025005 (2021).

4 Schoelkopf, R. J. & Girvin, S. M. Wiring up quantum systems. *Nature* **451**, 664–669 (2008).

5 Gu, X., Kockum, A. F., Miranowicz, A., Liu, Y.-x. & Nori, F. Microwave photonics with superconducting quantum circuits. *Phys. Rep.* **718-719**, 1–102 (2017).

6 Bozkurt, A. B., Golami, O., Yu, Y., Tian, H. & Mirhosseini, M. A mechanical quantum memory for microwave photons. *Nat. Phys.* **21**, 1469–1474 (2025).

7 Zhao, H., Chen, W. D., Kejriwal, A. & Mirhosseini, M. Quantum-enabled microwave-to-optical transduction via silicon nanomechanics. *Nat. Nanotechnol.* **20**, 602–608 (2025).

8 Lauk, N. *et al.* Perspectives on quantum transduction. *Quantum Sci. Technol.* **5**, 020501 (2020).

9 Degen, C. L., Reinhard, F. & Cappellaro, P. Quantum sensing. *Rev. Mod. Phys.* **89**, 035002 (2017).

10 Senanian, A. *et al.* Microwave signal processing using an analog quantum reservoir computer. *Nat. Commun.* **15**, 7490 (2024).

11 Yuan, J. *et al.* Quantum sensing of microwave electric fields based on rydberg atoms. *Rep. Prog. Phys.* **86**, 106001 (2023).

12 Hou, J. T. & Liu, L. Strong coupling between microwave photons and nanomagnet magnons. *Phys. Rev. Lett.* **123**, 107702 (2019).

13 Lenk, B., Ulrichs, H., Garbs, F. & Münzenberg, M. The building blocks of magnonics. *Phys. Rep.* **507**, 107–136 (2011).

14 Zhang, X., Zou, C. L., Jiang, L. & Tang, H. X. Strongly coupled magnons and cavity microwave photons. *Phys. Rev. Lett.* **113**, 156401 (2014).

15 Tabuchi, Y. *et al.* Hybridizing ferromagnetic magnons and microwave photons in the quantum limit. *Phys. Rev. Lett.* **113**, 083603 (2014).

16 Xu, D. *et al.* Quantum control of a single magnon in a macroscopic spin system. *Phys. Rev. Lett.* **130**, 193603 (2023).

17 Goryachev, M. *et al.* High-cooperativity cavity qed with magnons at microwave frequencies. *Phys. Rev. Appl.* **2**, 054002 (2014).

18 Mochizuki, M. Spin-wave modes and their intense excitation effects in skyrmion crystals. *Phys. Rev. Lett.* **108**, 017601 (2012).

19 Petrova, O. & Tchernyshyov, O. Spin waves in a skyrmion crystal. *Phys. Rev. B* **84**, 214433 (2011).

20 Cheng, Z., Chen, H., Huang, S. & Wu, Y. Spin wave mode coupling in antiskyrmions induced by rotational symmetry breaking. *Phys. Rev. B* **109**, 104431 (2024).

21 Okamura, Y. *et al.* Microwave magnetoelectric effect via skyrmion resonance modes in a helimagnetic multiferroic. *Nat. Commun.* **4**, 2391 (2013).

22 Satywali, B. *et al.* Microwave resonances of magnetic skyrmions in thin film multilayers. *Nat. Commun.* **12**, 1909 (2021).

23 Mühlbauer, S. *et al.* Skyrmion lattice in a chiral magnet. *Science* **323**, 915–919 (2009).

24 Nagaosa, N. & Tokura, Y. Topological properties and dynamics of magnetic skyrmions. *Nat. Nanotechnol.* **8**, 899–911 (2013).

25 Yu, X. Z. *et al.* Real-space observation of a two-dimensional skyrmion crystal. *Nature* **465**, 901–904 (2010).

26 Guslienko, K. Y. & Gareeva, Z. V. Gyrotropic skyrmion modes in ultrathin magnetic circular dots. *IEEE Magn. Lett.* **8**, 1–5 (2017).

27 Hoffmann, M. *et al.* Antiskyrmions stabilized at interfaces by anisotropic dzyaloshinskii-moriya interactions. *Nat. Commun.* **8**, 308 (2017).

28 Karube, K. *et al.* Room-temperature antiskyrmions and sawtooth surface textures in a non-centrosymmetric magnet with s(4) symmetry. *Nat. Mater.* **20**, 335–340 (2021).

29 Koshibae, W. & Nagaosa, N. Theory of antiskyrmions in magnets. *Nat. Commun.* **7**, 10542 (2016).

30 Nakamura, D. *et al.* Observation of resonant spin excitation with in-plane spin wave propagation in hybrid antiskyrmion strings. *Phys. Rev. B* **110**, 024418 (2024).

31 David B. Williams , C. B. C. *Transmission electron microscopy*. (Springer, 2009).

32 Feist, A. *et al.* Ultrafast transmission electron microscopy using a laser-driven field emitter: Femtosecond resolution with a high coherence electron beam. *Ultramicroscopy* **176**, 63–73 (2017).

33 Lobastov, V. A., Srinivasan, R. & Zewail, A. H. Four-dimensional ultrafast electron microscopy. *Proc. Natl. Acad. Sci. U.S.A.* **102**, 7069–7073 (2005).

34 Zewail, A. H. Four-dimensional electron microscopy. *Science* **328**, 187–193 (2010).

35 Liu, C. *et al.* Correlated spin-wave generation and domain-wall oscillation in a topologically textured magnetic film. *Nat. Mater.* **24**, 406–413 (2025).

36 Reisbick, S. A. *et al.* Stroboscopic ultrafast imaging using rf strip-lines in a commercial transmission electron microscope. *Ultramicroscopy* **235**, 113497 (2022).

37 Reisbick, S. A. *et al.* Characterization of transverse electron pulse trains using rf powered traveling wave metallic comb striplines. *Ultramicroscopy* **249**, 113733 (2023).

38 Guang, Y. *et al.* Confined antiskyrmion motion driven by electric current excitations. *Nat. Commun.* **15**, 7701 (2024).

39 Li, Z. *et al.* Topology-guided rotational dynamics of magnetic soliton assemblies under pulsed current. *Nat. Commun.* (2026).

40 Tang, J. *et al.* Sewing skyrmion and antiskyrmion by quadrupole of bloch points. *Sci. Bull.* **68**, 2919–2923 (2023).

41 Yasin, F. S. *et al.* Bloch point quadrupole constituting hybrid topological strings revealed with electron holographic vector field tomography. *Adv. Mater.* **36**, 2311737 (2024).

42 De Graef, M. & Zhu, Y. Quantitative noninterferometric lorentz microscopy. *J. Appl. Phys.* **89**, 7177–7179 (2001).

43 Leliaert, J. *et al.* Fast micromagnetic simulations on gpu—recent advances made with mumax$^3$. *J. Phys. D: Appl. Phys.* **51**, 123002 (2018).

44 Vansteenkiste, A. *et al.* The design and verification of mumax3. *AIP Adv.* **4**, 107133 (2014).

45 Jin, C. D. *et al.* Spin-wave modes of elliptical skyrmions in magnetic nanodots. *New J. Phys.* **24**, 043005 (2022).

46 Yang, J., Kim, J., Abert, C., Suess, D. & Kim, S.-K. Stability of skyrmion formation and its abnormal dynamic modes in magnetic nanotubes. *Phys. Rev. B* **102**, 094439 (2020).

47 Leliaert, J. *et al.* Adaptively time stepping the stochastic landau-lifshitz-gilbert equation at nonzero temperature: Implementation and validation in mumax3. *AIP Adv.* **7**, 125010 (2017).

48 Karube, K. *et al.* Doping control of magnetic anisotropy for stable antiskyrmion formation in schreibersite (fe,ni)3p with s4 symmetry. *Adv. Mater.* **34**, 2108770 (2022).

49 Yasin, F. S. *et al.* Heat current-driven topological spin texture transformations and helical q-vector switching. *Nat. Commun.* **14**, 7094 (2023).

## Acknowledgements

This work was supported in part by the U.S. Department of Energy (DOE), Basic Energy Sciences, Materials Science and Engineering Division, under Contract DESC0012704, received by Y.M. Z., Japan Science and Technology Agency (JST) CREST program (Grant No. JPMJCR20T1, received by X.Z. Y. and Y. Taguchi) and FOREST (Grant No. JPMJFR235R, received by K. K.), the Japan Society for the Promotion of Science (JSPS) Grants-in-Aid for Scientific Research (Grant No. 23K26534 received by K. K., Grant No. 23H05431 received by Y. Tokura, Grant No. 23K03291 received by W. Koshibae, Grant Nos. 24H00197, 24H02231 and 24K00583 received by N. N.), and the RIKEN TRIP initiative received by X.Z. Y., Y. Taguchi and N. N.

## Author Contributions

Y.M. Z. and X.Z. Y. conceived the project. Z.L. L. prepared the samples for experiments. D. N., S. R. and X. Z. Y. performed the L-TEM experiments with the support of C.H. L. and M.G. H. Z.L. L. and X.Z. Y. analyzed the results and wrote the manuscript with assistance from Y. Tokura. W. K. and N. N. supported the theoretical part. Z.L. L. conducted micromagnetic simulations. K. K. and Y. Taguchi prepared the FNPP bulk samples. All authors discussed the data and commented on the manuscript.

## Competing Interests

The authors declare no competing interests.

## Extended data

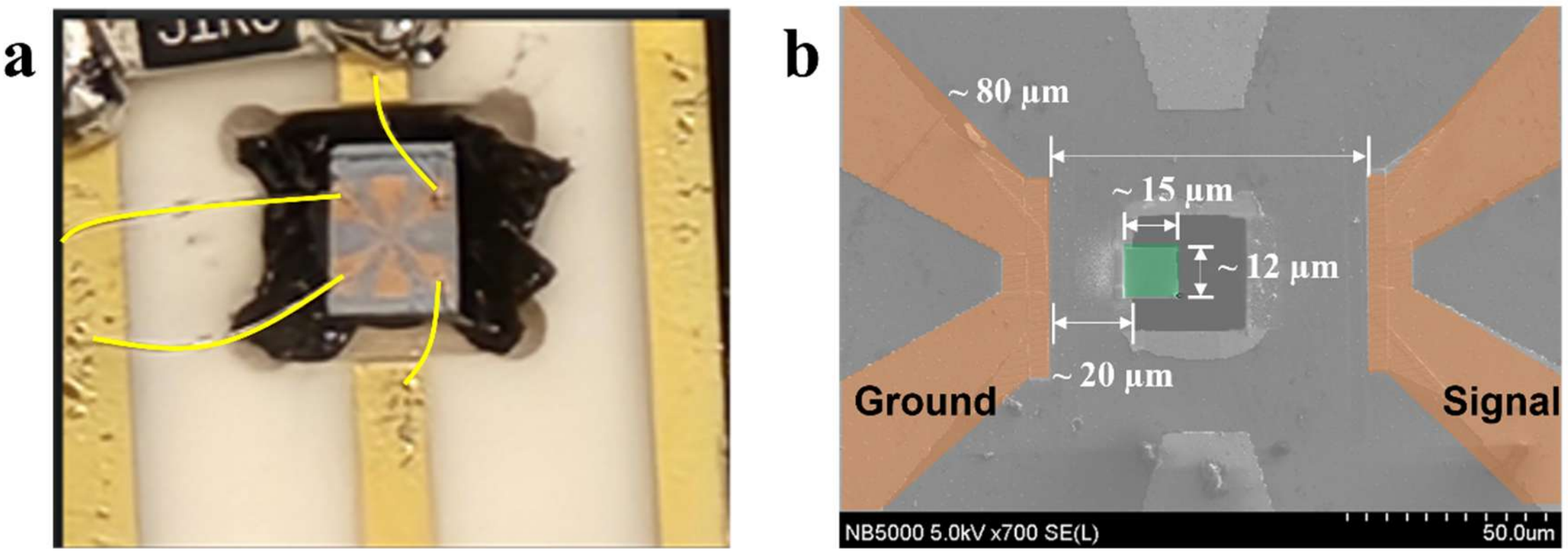

**Extended Data Fig. 1 | Electrical connection and SEM image of the microwave-compatible chip. a,** Optical image of the chip mounted on the tip of the microwave-compatible TEM holder. The wide Au pads on the left side of the holder tip are wire-bonded to the left chip electrodes and serve as the ground terminal. The central wide Au pads are wire-bonded to the right chip electrodes and serve as the signal terminal. **b,** SEM image of the chip with the FIB-prepared FNPP sample placed across the central hole. The specimen dimensions and electrode spacing are indicated. Consistent with **a**, the left electrodes function as the ground terminal and the right electrodes serve as the signal terminal.

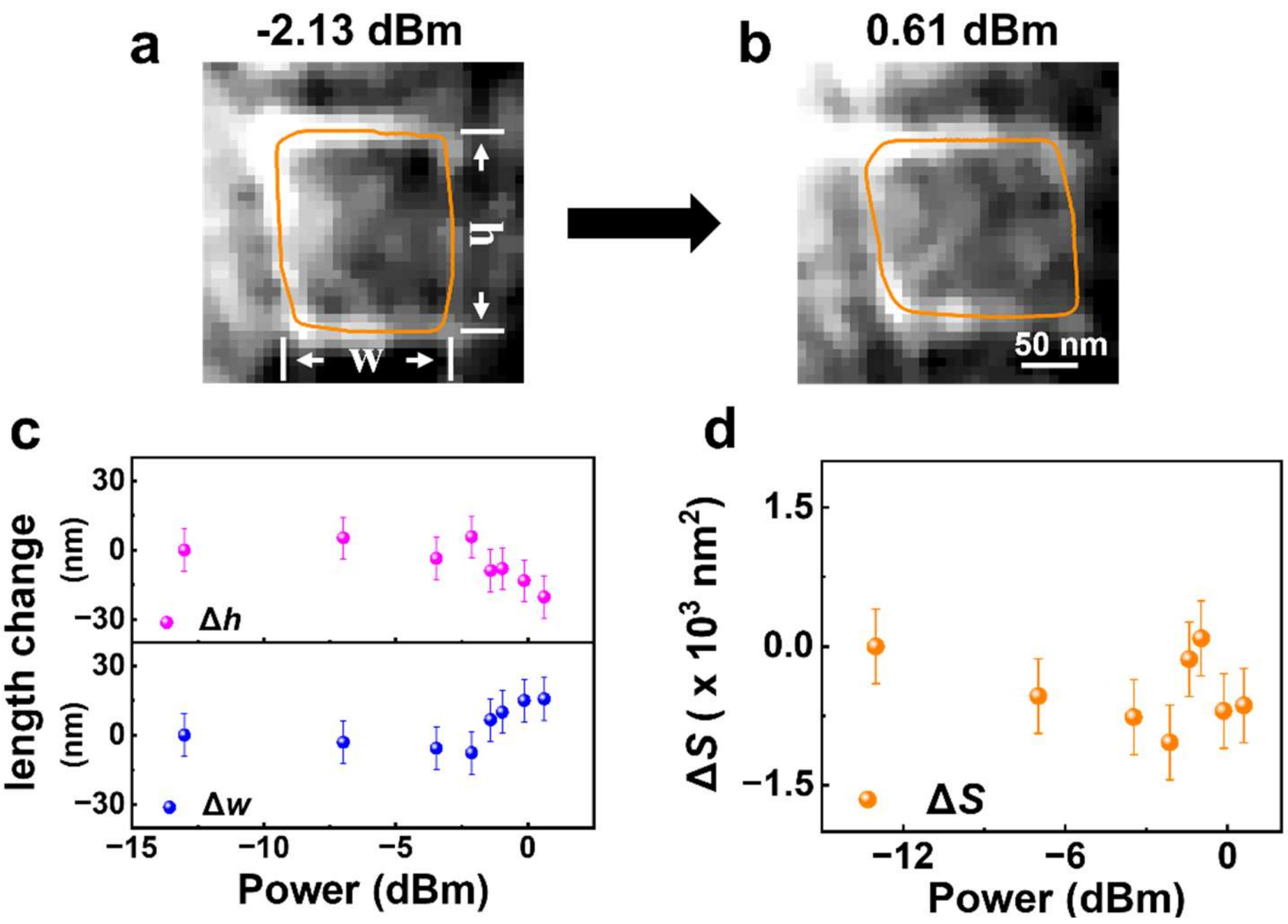

**Extended Data Fig. 2 | Power-dependent deformation of an antiskyrmion under microwave excitation in Fresnel mode of L-TEM. a-b,** L-TEM images of the same antiskyrmion recorded at two representative microwave powers under a fixed excitation frequency of 4 GHz. Yellow outlines mark the antiskyrmion boundaries extracted using the same image-processing procedure used in the main text. The scale bar is 50 nm. **c,** Relative changes in the antiskyrmion width and height as a function of microwave power, referenced to the low-power state of -13 dBm. The antiskyrmion shows anisotropic deformation, characterized by expansion along one direction and compression along the orthogonal direction, with the threshold power above −2.13 dBm. **d,** Relative change in the enclosed area as a function of microwave power. Error bars represent one standard deviationof the experimental data. The absence of the inner vortex-like contrast is attributed to the metastable character of the surface-skyrmion structure responsible for this contrast, which is therefore not present in every antiskyrmion in the specimen.

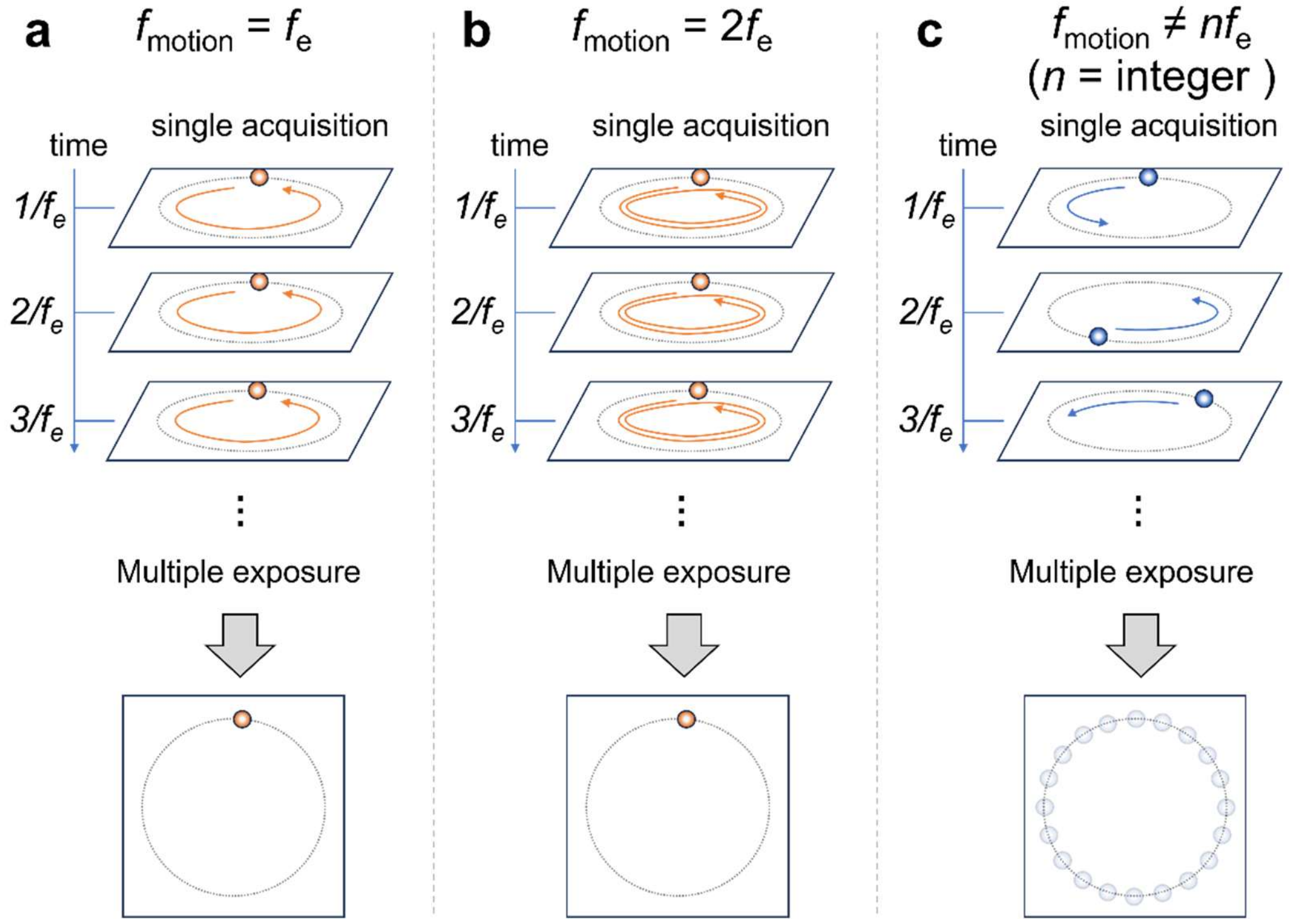

**Extended Data Fig. 3 | Schematic illustration of the intrinsic frequency selectivity in phase-locked UEM measurements.** Schematic illustration of single-frame formation by accumulating multiple pulsed electron-beam exposures with repetition frequency $f_e$. The object undergoes counterclockwise circular motion with the frequency of $f_{motion}$. **a,** Frequency-matched case, $f_{motion} = f_e$. The object is sampled at the same phase at each exposure and remains localized in the accumulated frame. **b,** Harmonic case, $f_{motion} = 2f_e$. Sampling again occurs at a fixed phase, preserving a well-defined object position. **c,** Non-synchronized case. When $f_{motion}$ is not synchronized with $f_e$ or its integer harmonics, repeated exposures sample different phases, causing the accumulated signal to spread along the trajectory and become averaged out. Phase-locked UEM therefore selectively preserves dynamical components synchronized with $f_e$ and its integer harmonics.

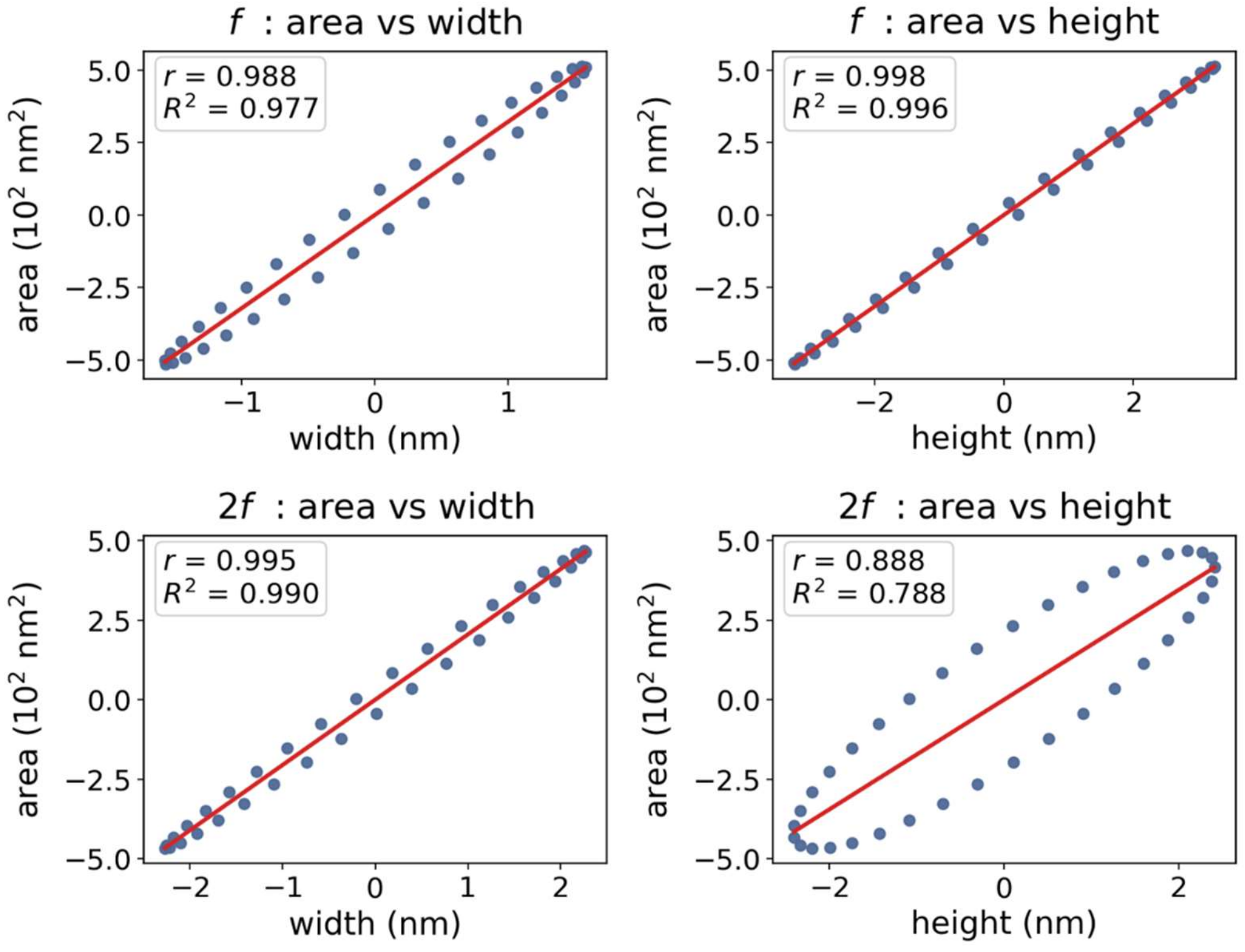

**Extended Data Fig. 4 | Correlation of antiskyrmion area change with width and height modulation. a-b,** Correlations between the fundamental-frequency (4 GHz) component of the antiskyrmion area change and the corresponding width (**a**) and height (**b**) modulations. The area variation is more strongly correlated with the height modulation at the fundamental frequency. **c,d,** Correlations between the second-harmonic (8 GHz) component of the antiskyrmion area change and the corresponding width (**c**) and height (**d**) modulations. The second-harmonic area response shows a stronger correlation with the width modulation. Blue symbols represent the extracted data points from the bandpass-filtered results, and red lines indicate linear fits. The Pearson correlation coefficient, $r$, is determined by $r = \sum(x_i - \overline{x})(y_i - \overline{y}) / \sqrt{(\sum(x_i - \overline{x})^2 \sum(y_i - \overline{y})^2)}$ and the coefficient of determination from the linear regression, $R^2$, is determined as $R^2 = 1 - \sum(y_i - y_i')^2 / \sum(y_i - \overline{y})^2$, where $y_i'$ denotes the fitted value from the linear regression. Both parameters describe the linear correlation between the two variables, with values close to 1 indicating a strong positive linear correlation.

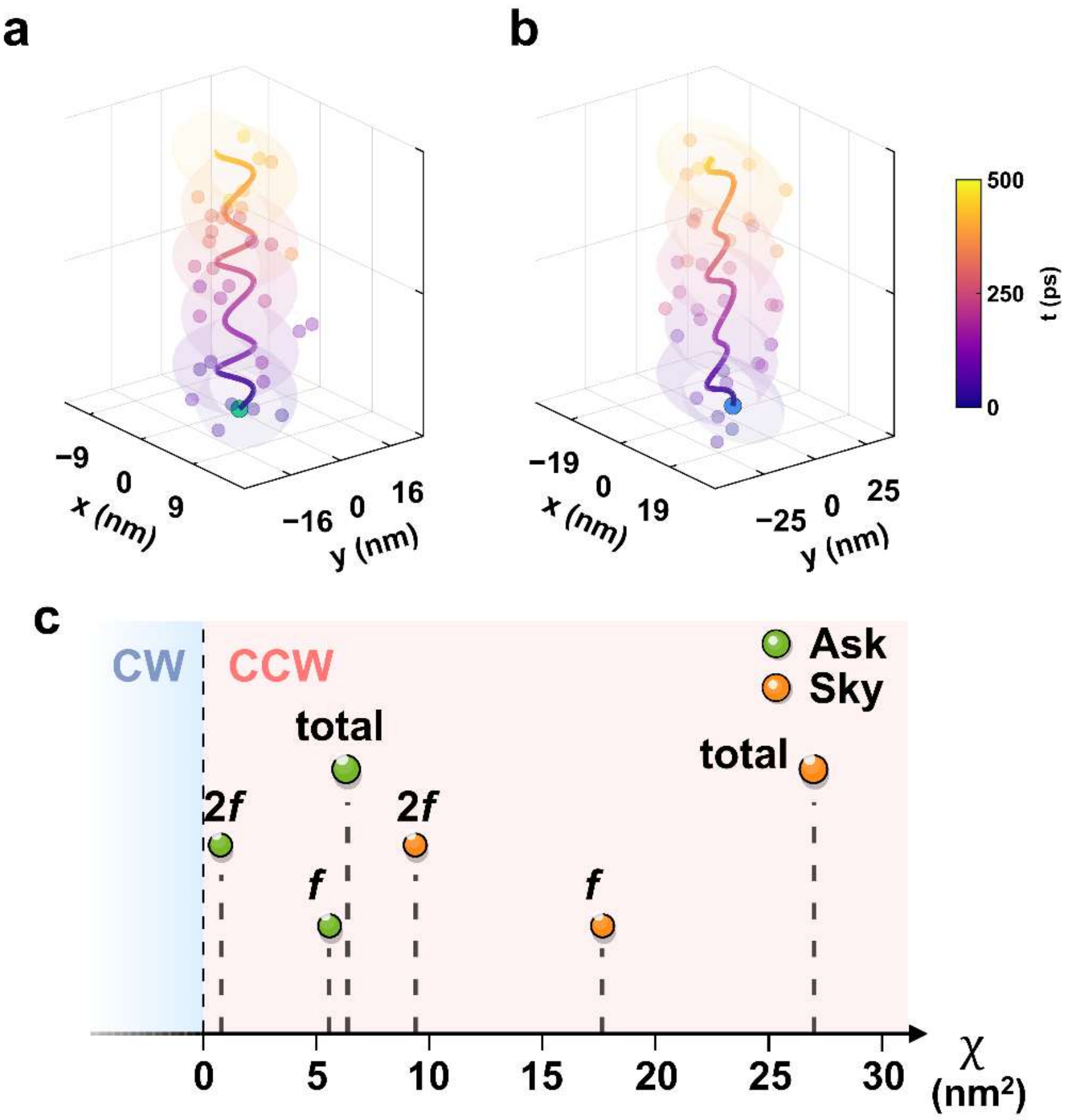

**Extended Data Fig. 5 | Time-resolved core trajectories under 5.25 GHz microwave excitation.** **a–b,** The temporal evolution of the extracted core position (spots) of the antiskyrmion (**a**) and surface skyrmions (**b**) respectively, together with the corresponding bandpass-filtered trajectories. The shaded ribbons denote the one-standard-deviation envelopes derived from the residual covariance of the raw core positions around the fitted trajectory. The rigid translational shift observed in Supplementary Video 3 was subtracted before trajectory reconstruction. **c**, Calculated signed area $\chi$ for the bandpass-filtered core trajectories of the antiskyrmion and surface skyrmion. The total reconstructed motion as well as the fundamental and second-harmonic components are evaluated separately. Both cores retain counterclockwise rotational motion, with larger rotational amplitudes than those observed under 4 GHz excitation.

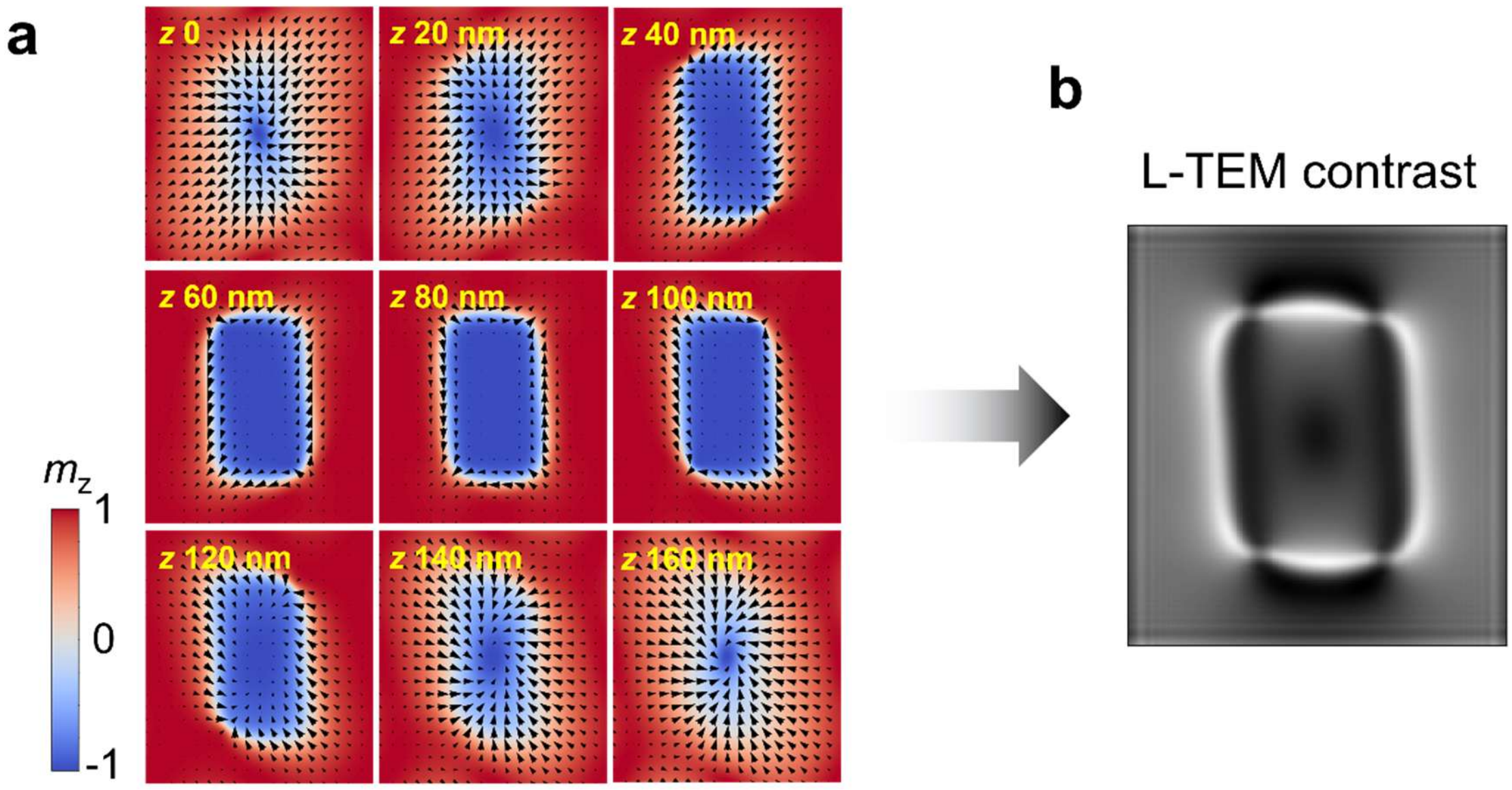

**Extended Data Fig. 6 | Simulated *z*-dependent spin configurations and L-TEM contrast of the hybrid antiskyrmion texture. a,** Representative spin configurations extracted from selected slices at different z positions of the three-dimensional micromagnetic model. The color denotes the out-of-plane magnetization component, $m_z$, and arrows indicate the in-plane magnetization direction. The hybrid texture consists of skyrmions near the surfaces and the central antiskyrmion in the interior. **b**, Corresponding simulated L-TEM contrast of the total hybrid structure. The central vortex-like contrast originates from the swirling core of the surface skyrmions.

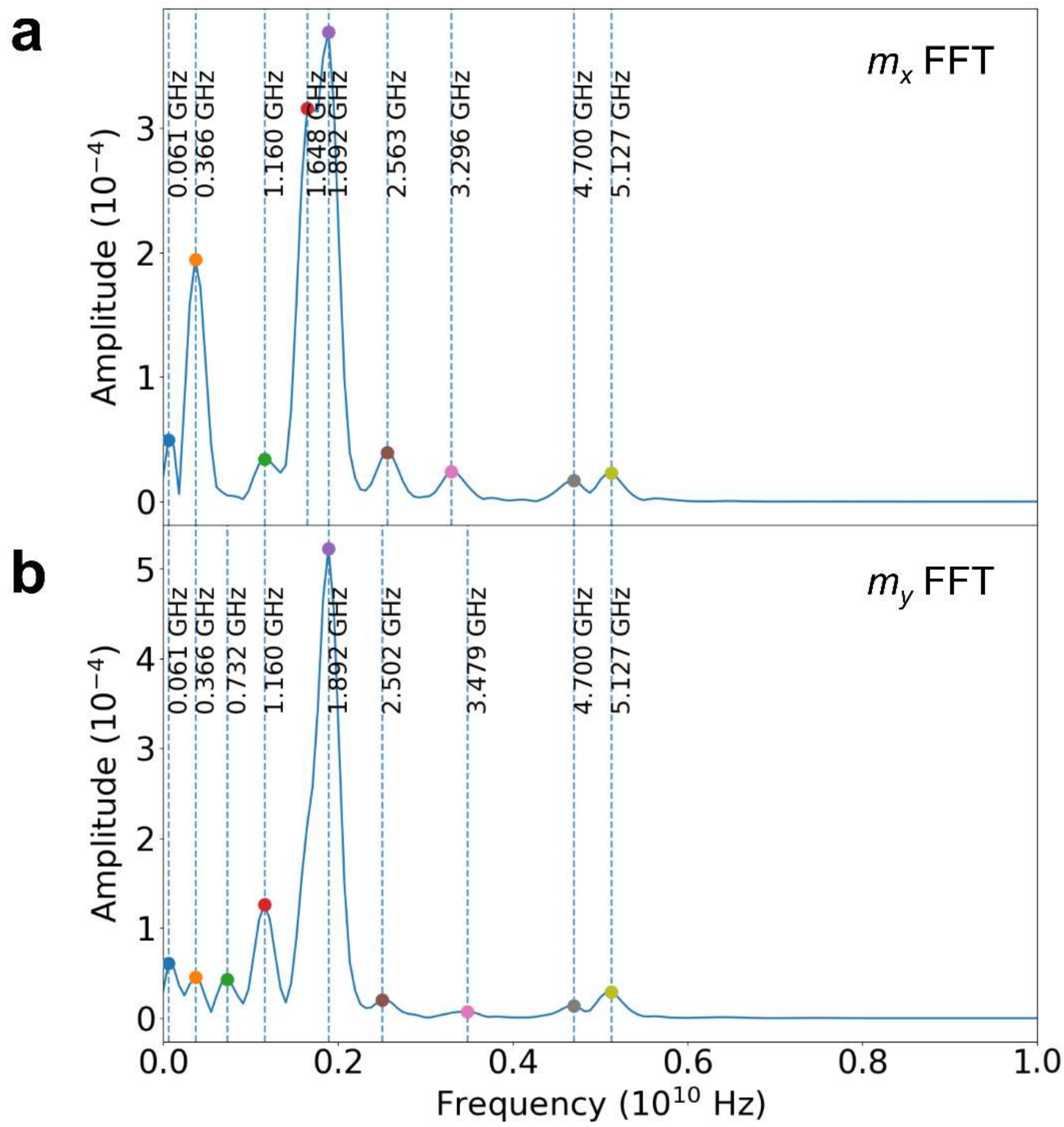

**Extended Data Fig. 7 | Simulated spectrum of the hybrid antiskyrmion under sinc-pulse excitation.** Fourier spectra of the dynamical response obtained from the temporal evolution of the magnetization components after excitation by a sinc-shaped magnetic-field pulse with both in-plane and out-of-plane components, $B_{\mathrm{sinc}}(t) = B_{\mathrm{amp}}\sin[\pi f_0(t-t_0)]/[\pi f_0(t-t_0)]$, with $B_{\mathrm{amp}} = 7\mathrm{mT}$ and $f_0 = 50\mathrm{GHz}$. This sinc excitation provides broad spectral content from 0 to 25 GHz and thus enables simultaneous excitation of multiple dynamical modes. **a**, FFT spectrum of $m_x$. **b**, FFT spectrum of $m_y$. The dashed vertical lines mark the peak frequencies extracted from the spectra. Resonance peaks at 4.70 and 5.13 GHz are observed, close to the experimentally applied frequencies of 4 and 5.25 GHz.

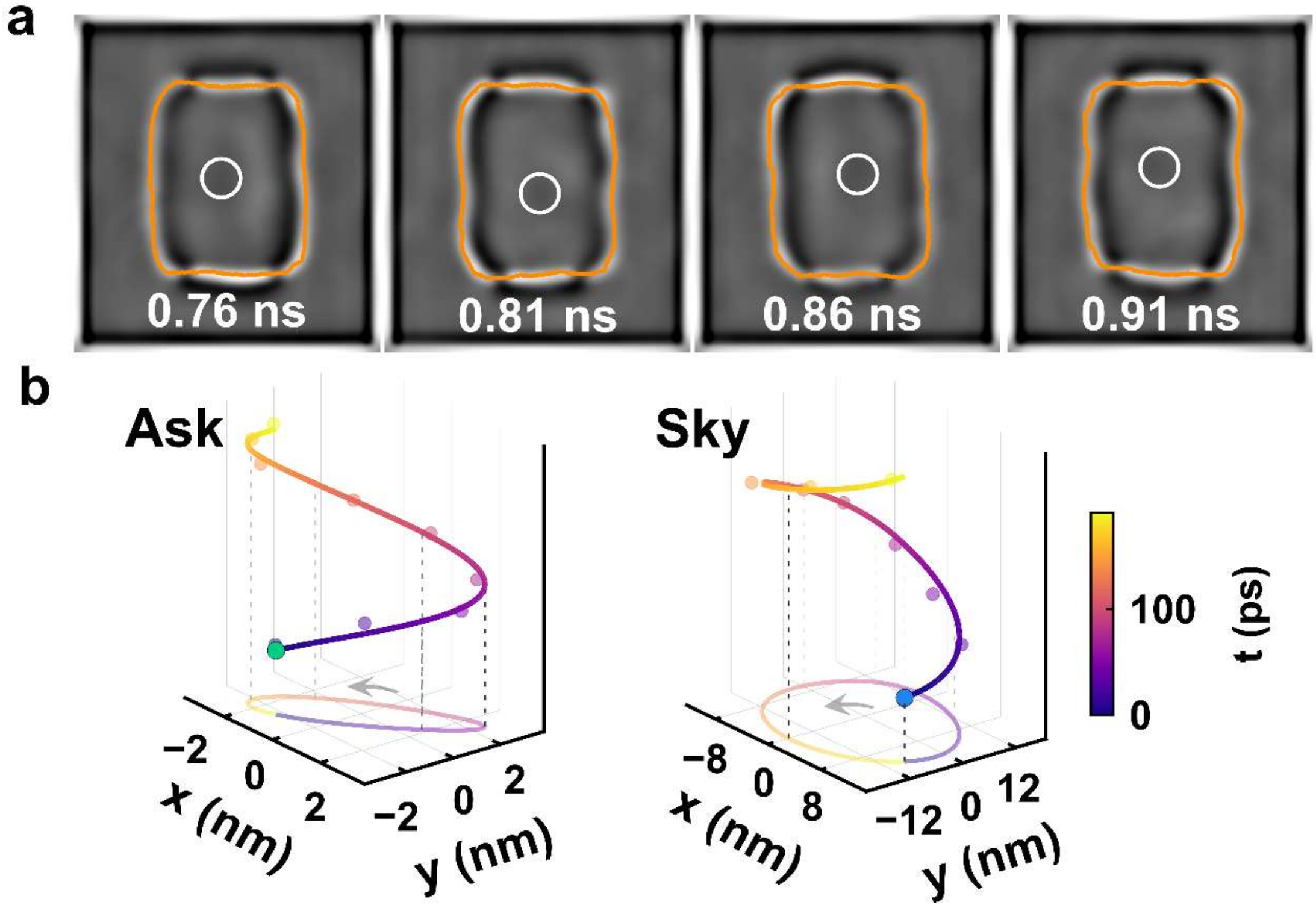

**Extended Data Fig. 8 | Simulated microwave-driven dynamics of the hybrid antiskyrmion at 5.25 GHz. a,** Simulated L-TEM contrast images at representative delay times over one microwave period under a 5.25 GHz sinusoidal magnetic field. The yellow outline marks the extracted antiskyrmion contour, and the white dashed circle marks the vortex-like contrast associated with the surface skyrmions. **b**, The temporal evolution of the extracted core position (spots) of the antiskyrmion and surface skyrmions respectively, together with the corresponding bandpass-filtered trajectories. The results show that the rotation is counterclockwise and has a larger amplitude than that obtained under 4 GHz excitation in Fig. 4**b-e**, consistent with the experimental trend. The color bar encodes the time.

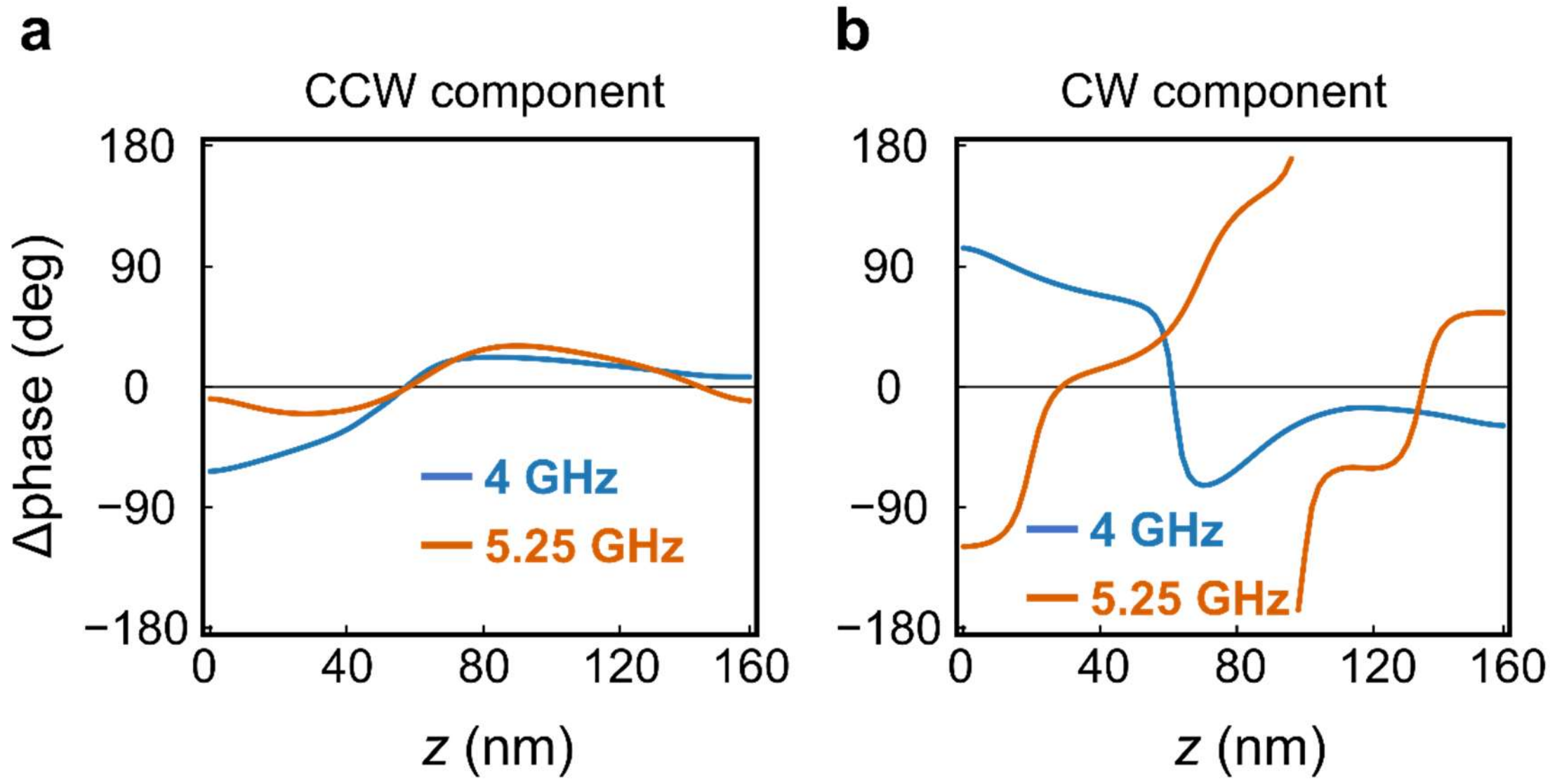

**Extended Data Fig. 9 | *z*-dependent phase coherence of CW and CCW core-trajectory components.** *z*-dependent phase differences between the CW or CCW components of the simulated core trajectories in individual *x*-*y* slices and those of the L-TEM-projected inner vortex-like core trajectory. Blue and orange curves denote 4 and 5.25 GHz excitation, respectively. Phases are wrapped to [-180°, 180°], with line breaks indicating wrapping discontinuities. **a,** CCW component. The phase differences remain close to zero with little variation along *z*, indicating coherent CCW motion that adds constructively in projection. **b,** CW component. The phase differences vary strongly along *z*, indicating dephased CW motion that is partially cancelled or suppressed in the projected L-TEM signal.